# Field-tuning of ultrafast magnetization fluctuations in $Sm_{0.7}Er_{0.3}FeO_3$

M. A. Weiss[1,*], J. Schlegel[1], D. Anić[1], E. Steiner[1], F. S. Herbst[1], M. Nakajima[2], T. Kurihara[1,3], A. Leitenstorfer[1], U. Nowak[1], & S. T. B. Goennenwein[1]

[1]Department of Physics, University of Konstanz, D-78457 Konstanz, Germany
[2]Institute of Laser Engineering, Osaka University, 565-0871 Osaka, Japan
[3]Department of Basic Science, The University of Tokyo, 153-8902 Tokyo, Japan

The properties of spin fluctuations in antiferromagnets are largely unexplored, in particular at ultrafast timescales. Here, we employ femtosecond noise correlation spectroscopy (FemNoC) to experimentally study magnetization fluctuations in the canted antiferromagnet $Sm_{0.7}Er_{0.3}FeO_3$ across its spin reorientation transition and under external magnetic fields. By comparing our measurements to atomistic spin noise and Monte Carlo simulations, we find that the amplitude of the spin noise is governed by the free energy, with stronger fluctuations in regions where the potential landscape softens. We furthermore demonstrate that external magnetic fields suppress spin fluctuations and enhance the quasi-ferromagnetic magnon frequency by effectively stiffening the potential. These results highlight an effective route for tuning ultrafast magnetization fluctuations via external parameters.

## I. INTRODUCTION

The development of high-speed, energy-efficient spintronic devices relies on materials that exhibit high-frequency and low-noise magnetization dynamics. Antiferromagnets (AFMs) are particularly promising in this regard due to their terahertz-range magnon frequencies, absence of stray fields, and resilience to external perturbations [1–4]. While much research has focused on the deterministic control of magnons in AFMs [5–7], their intrinsic spin fluctuation dynamics, remain largely unexplored. Gaining insights into these fluctuations is crucial for designing a new generation of devices with minimized dissipation.

Recently, femtosecond noise correlation spectroscopy (FemNoC) [8–10] has emerged as a powerful technique for probing high-frequency spin fluctuations. By measuring correlations in polarization noise imprinted on pairs of consecutive femtosecond probe pulses via the Faraday effect, FemNoC provides direct access to the time-domain dynamics of magnon fluctuations.

In this work, we employ FemNoC to investigate the canted antiferromagnet $Sm_{0.7}Er_{0.3}FeO_3$, focusing on the interplay between magnon fluctuations, the spin reorientation transition (SRT), and the free energy potential landscape. By systematically varying temperature and external magnetic field, we reveal how changes in the potential landscape dictate the fluctuation amplitude of the quasi-ferromagnetic (qF) magnon mode. Our findings show that thermal fluctuations are enhanced in regions where the free energy potential softens, a trend supported by atomistic spin noise and Monte Carlo simulations. Additionally, we demonstrate that an external magnetic field suppresses fluctuations and increases the qF magnon frequency by effectively stiffening the potential, offering a direct means of tuning ultrafast magnetization fluctuations via external control parameters.

## II. METHODS

### A. Experimental sample

The sample used in this study is a 10 μm thick single crystal of the orthoferrite $Sm_{0.7}Er_{0.3}FeO_3$ [9–12] prepared via the floating zone technique [13–15]. In this material, $Fe^{3+}$ electron spins are antiferromagnetically coupled, with a slight spin canting induced by the Dzyaloshinskii–Moriya interaction [16,17], resulting in a weak net ferromagnetic magnetization **M**. Orthoferrites support two exchange modes at multi-THz frequencies and two magnon modes in the sub-THz regime: the quasi-ferromagnetic (qF) and quasi-antiferromagnetic (qAF) modes [18,19]. The qF mode corresponds to a precession of the weak ferromagnetic moment around its equilibrium axis, while the qAF mode involves longitudinal oscillations of this moment [9].

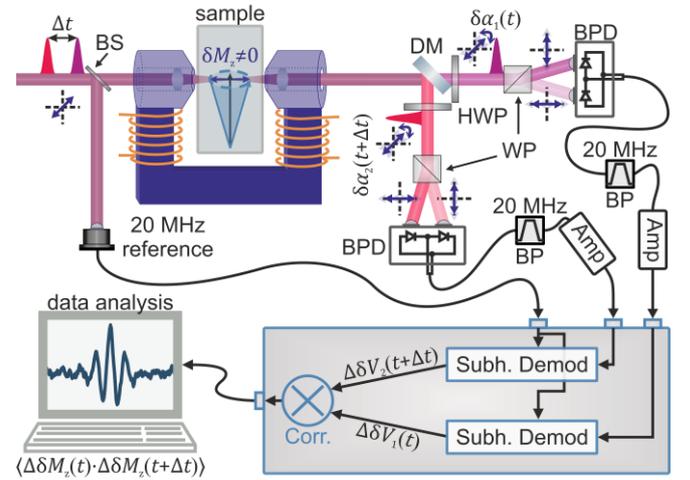

FIG 1: Experimental setup and electronic data processing. The working principle is discussed in the main text. DM: Dichroic mirror; HWP: Half-wave plate; WP: Wollaston prism; BPD: Balanced photodetector; BP: electronic bandpass filter; Amp: Amplifier; Subh. Demod: Subharmonic demodulation; Corr: Real-time correlation.

$Sm_{0.7}Er_{0.3}FeO_3$ undergoes a temperature-induced spin reorientation transition (SRT) near room temperature [9,11], where the magnetic anisotropy changes while the antiferromagnetic spin order remains intact. Within the SRT region ($T_L < T < T_U$), **M** continuously rotates from the crystallographic $a$-axis ($\Gamma_2$ configuration) at the lower critical temperature $T_L$ to the $c$-axis at the upper critical temperature $T_U$ ($\Gamma_4$ configuration) [20].

*Contact author: marvin.weiss@uni-konstanz.de

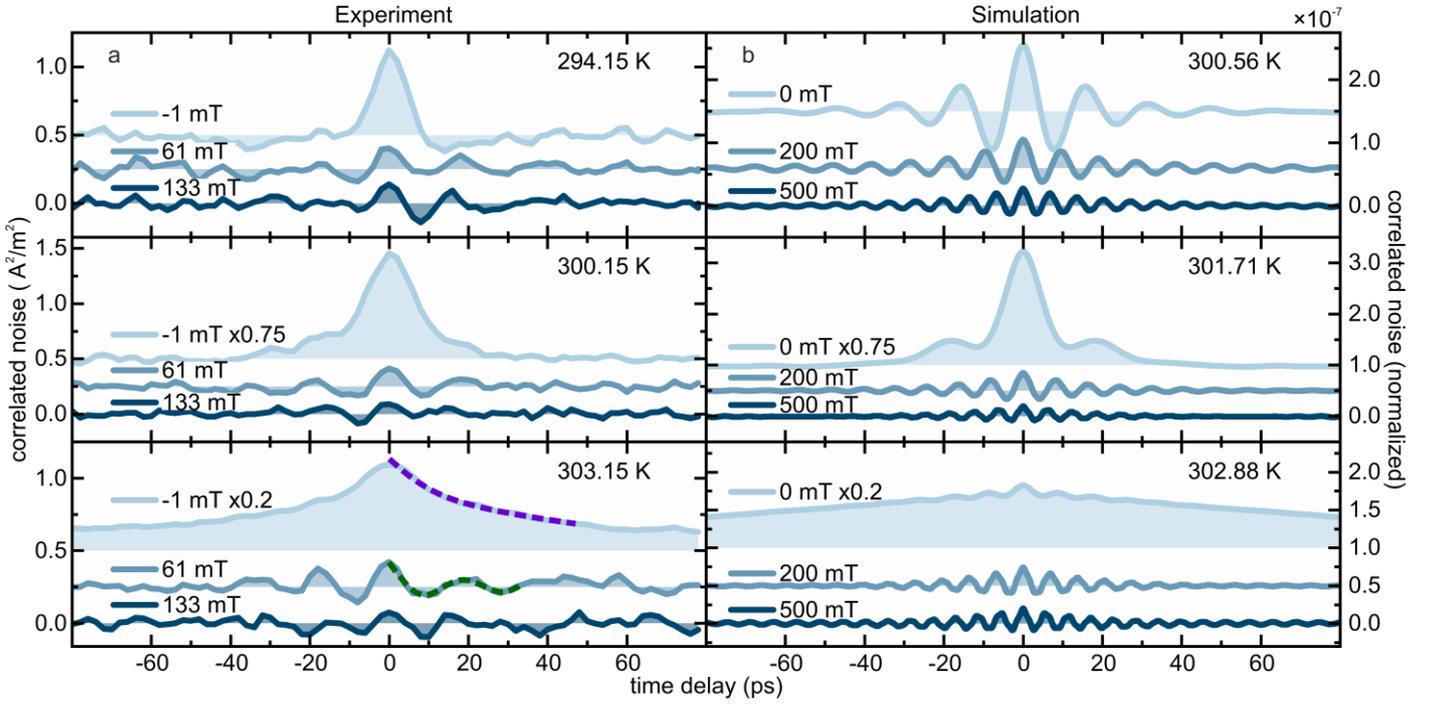

FIG 2: Spin noise dynamics in $Sm_{0.7}Er_{0.3}FeO_3$ as a function of probe-probe time delay $\Delta t$. **a** Experimental correlation time traces obtained from temperature scans in the range of 293 K – 325 K and under various out-of-plane magnetic fields. Background subtraction was performed as described in Appendix A. The purple and green dotted lines are guide to the eyes, representing the exponential decay and damped picosecond oscillation around the zero-crossing line corresponding to ultrafast random telegraph noise, and quasi-ferromagnetic magnon noise, respectively. **b** Simulated noise correlation time traces for different magnetic fields along the crystalline $c$-axis and temperatures within the SRT region. For clarity, both experimental and simulated waveforms are vertically shifted, with the shaded area indicating the zero-line. Some traces additionally are scaled for better comparison.

Phenomenologically, the $a$-$c$ plane spin reorientation of the net magnetization is understood in terms of the softening of a Landau-type free energy [7,21–24]:

$$F(\theta) = F_0 + K_2 \cos^2(\theta) + K_4 \cos^4(\theta) - \frac{1}{\mu_0} \mathbf{MB} \quad (1)$$

Here, $\theta$ is the angle between the net magnetization $\mathbf{M}$ and the $a$-axis, $F_0$ is a constant free energy term, $K_2$ and $K_4$ are the anisotropy constants of second and fourth order, respectively, $\mu_0$ is the vacuum permeability and $\mathbf{B}$ is the external magnetic field. $K_4$ is approximately constant, whereas $K_2$ is assumed to be temperature-dependent to account for the temperature-induced spin reorientation [21]:

$$K_2(T) = 2K_4 \frac{T - T_U}{T_U - T_L} \quad (2)$$

Minimization of Eq. (1) yields the real and stable solutions $\theta_1 = 0$, $\theta_2 = \frac{\pi}{2}$, and $\theta_3 = \arccos\left(\sqrt{-\frac{K_2}{2K_4}}\right)$ for temperatures $T < T_L$, $T > T_U$, and $T_L < T < T_U$, respectively [21], thereby capturing the $a$-axis to $c$-axis SRT in $Sm_{0.7}Er_{0.3}FeO_3$.

Furthermore, the vanishing anisotropy difference between the $a$- and $c$-axes at $T_L$ leads to an increased magnetic susceptibility in the $a$-$c$ plane [25] and significant softening of the qF mode resonance frequency [9,11,22,26,27]. This softening is accompanied by a strong enhancement of the qF magnon fluctuation amplitude and the emergence of picosecond stochastic switching (random telegraph noise, RTN) [9].

### B. FemNoC technique

In this work, we use femtosecond noise correlation spectroscopy [9,10,28–30] (FemNoC) to measure spontaneous magnetization fluctuations in $Sm_{0.7}Er_{0.3}FeO_3$. A schematic of our experimental setup is shown in FIG 1. FemNoC operates as follows: Two linearly polarized femtosecond laser pulse trains (40 MHz repetition rate) with slightly different center wavelengths (767 nm, purple; 775 nm red) and variable time delay $\Delta t$ are focused through the magnetic sample with a transmissive objective lens (numerical aperture NA = 0.4, working distance WD = 3.9 mm). The sample is positioned between the pole shoes of a custom-built electromagnet, which provides fields up to $\pm 450$ mT along the laser propagation. The magnet design allows to accommodate the objective lens within its inner structure and for the laser to transmit through its center, ensuring full compatibility with the optical setup.

Both probe pulse trains are polarized with the mean polarization angle $\langle \alpha_{1,2} \rangle = 45°$ with respect to the optical table. Due to the Faraday effect, the light probes experience additional polarization rotations $\delta\alpha_1$ and $\delta\alpha_2$ proportional to the out-of-plane spin fluctuations $\delta M_z$. In our experiment, the laser propagation direction is aligned with the crystalline $c$-axis of the sample, such that $\delta M_z = \delta M_c$. The polarization rotations $\delta\alpha_1$ and $\delta\alpha_2$ are measured in separate detection branches, each containing a half-wave plate (HWP), Wollaston prism (WP), and balanced photodetector (BPD). The WP splits the light into s- and p-polarized components (blue arrows), which the BPD converts into a voltage signal proportional to the optical power difference of the components. Consequently, if $\langle \alpha_{1,2} \rangle = 45°$, the BPDs output a voltage proportional to the polarization fluctuations $\delta\alpha_{1,2}$. The HWP is used to compensate for any static contributions to the Faraday rotation, e.g., induced by the static out-of-plane components of the net magnetization $\boldsymbol{M}$, and ensure $\langle \alpha_{1,2} \rangle \approx 45°$. This balancing condition yields a zero output voltage from the BPDs, rendering the system highly sensitive to slight polarization changes that produce a measurable finite signal. As a result, the HWP angles under which the BPDs are balanced contain information about $\boldsymbol{M}$, and can be used to determine the angle $\theta$ between $\boldsymbol{M}$ and the crystalline a-axis. This procedure is described in detail in Appendix C.

*Contact author: marvin.weiss@uni-konstanz.de

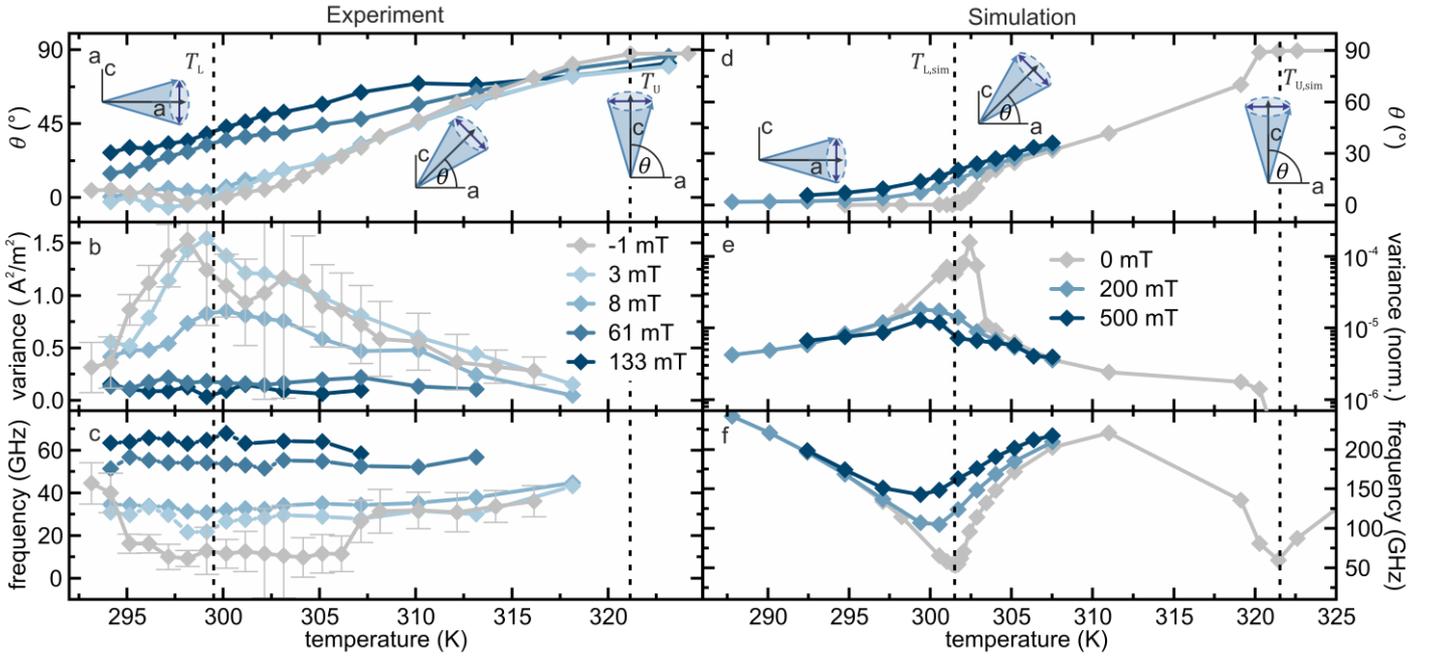

FIG 3: Temperature dependence of c-axis component of the static magnetization and quasi-ferromagnetic (qF) magnon noise dynamics in Sm$_{0.7}$Er$_{0.3}$FeO$_3$ under different c-axis magnetic fields. **a** Experimentally determined net magnetization angle $\theta$ as a function of temperature. The insets illustrate net magnetization direction (black arrow) and transverse fluctuations (blue arrow) relative to the crystalline a- and c-axes. $T_L$ and $T_U$ denote the lower and upper critical temperatures of the spin reorientation transition, respectively. **b,c** Variance and center frequency of the experimentally measured qF magnon fluctuations as functions of temperature and c-axis magnetic field. **d** Simulated net magnetization angle $\theta$ as a function of temperature. **e,f** Variance and center frequency of the simulated qF magnon fluctuations as functions of temperature and external magnetic field along the c-axis. Note that the $\theta$ curves for $B = 3$ mT and $B = 61$ mT are taken from reference [12].

To evaluate the fluctuation part of the magnetization, the BPD output voltage is filtered with a bandpass, amplified and subharmonically demodulated [10,31] in a lock-in amplifier (UHFLI, Zurich Instruments) using the 20 MHz reference of the laser pulse train. The resulting pulse-to-pulse voltage fluctuations $\Delta\delta V_{1,2}$ are multiplied in real time and averaged over approximately $10^8$ pulse pairs for each time delay $\Delta t$. This process yields the correlation trace of the out-of-plane magnetization dynamics $\langle \Delta\delta M_z(t)\Delta\delta M_z(t + \Delta t)\rangle$. Further details on the experimental setup and data processing scheme can be found in ref. [10].

### C. Atomistic spin model simulations

Our experimental findings are compared with simulations of the spin noise around the SRT in Sm$_{0.7}$Er$_{0.3}$FeO$_3$, resting on an atomistic spin model and the stochastic Landau-Lifshitz-Gilbert equation [32–34]. Sm$_{0.7}$Er$_{0.3}$FeO$_3$ is modelled as a generic orthoferrite with magnetic moments on the Fe sites only. The extended Heisenberg Hamiltonian for the to the unity normalized spins $\boldsymbol{S}^i$ (on lattice site $i$) reads

$$H = -\sum_{i,j} \boldsymbol{S}^i \mathfrak{J}^{ij} \boldsymbol{S}^i - \sum_i \boldsymbol{S}^i \kappa \boldsymbol{S}^i - \sum_{\substack{i \\ \eta,\nu \in \{x,y,z\}}} L_{\eta\nu}(S^i_\eta)^2 (S^i_\nu)^2 - \mu_S \boldsymbol{B} \sum_i \boldsymbol{S}^i.$$

(3)

The matrix $\mathfrak{J}^{ij}$ in the first term includes the isotropic Heisenberg exchange, the Dzyaloshinksii-Moriya interaction and a two-ion anisotropy, all taken into account up to the next-nearest neighbors. The second term is a second-order onsite anisotropy, competing with the aforementioned two-ion anisotropy, which leads to the SRT. The third term is a fourth-order onsite anisotropy, which determines the width of the SRT. The exact parameters can be found in ref. [9]. The last term is the Zeeman term with the magnetic field $\boldsymbol{B}$ and the amplitude of the spins $\mu_S = 3.66\, \mu_B$.

The dynamics of the spins is governed by the Landau-Lifshitz-Gilbert equation,

$$\dot{\boldsymbol{S}}^i = -\frac{\gamma}{\mu_S(1+\alpha^2)} \boldsymbol{S}^i \times (\boldsymbol{H}^i + \alpha \boldsymbol{S}^i \times \boldsymbol{H}^i),$$

(4)

with the gyromagnetic $\gamma$, here set to the one of a free electron, and the Gilbert damping parameter $\alpha = 0.0002$. The effective field is given by

$$\boldsymbol{H}^i = -\frac{\partial H}{\partial \boldsymbol{S}^i} + \boldsymbol{\zeta}^i,$$

(5)

with a stochastic Gaussian white noise uncorrelated in space and time,

*Contact author: marvin.weiss@uni-konstanz.de

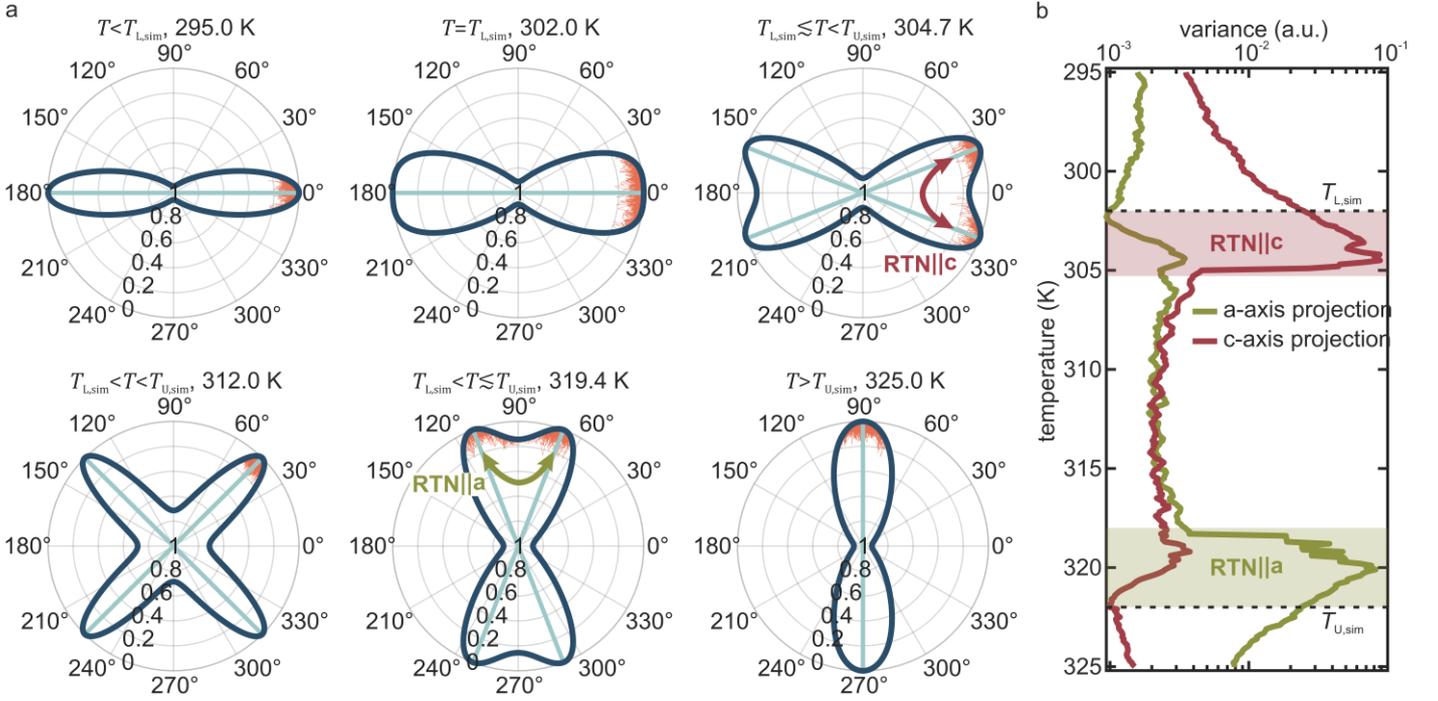

FIG 4: Monte Carlo simulated magnetization noise across the SRT region in $Sm_{0.7}Er_{0.3}FeO_3$. **a** Normalized free energy (dark blue) and Monte Carlo simulated net magnetization trajectories (orange) as functions of net magnetization angle $\theta$ and temperature. Light blue lines mark the easy-axis directions where the free energy is minimized. **b** Variances of the net magnetization projections along the $a$-axis (green) and $c$-axis (red) as a function of temperature. Green and red shaded regions indicate temperature ranges where random telegraph noise (RTN) occurs in the Monte Carlo simulation.

$$\langle \zeta^i \rangle = 0,$$

$$\langle \zeta_\eta^i(t)\zeta_\nu^j(t') \rangle = \frac{2\alpha k_B T \mu_S}{\gamma} \delta_{ij}\delta_{\eta\nu}\delta(t-t').$$

(6)

The time evolution of a system of $192^3$ spins is numerically calculated via the stochastic Heun's method.

### D. Monte Carlo simulation

To phenomenologically model magnetization noise in $Sm_{0.7}E_{0.3}FeO_3$ within the SRT regime, we perform Monte Carlo (MC) simulations of a magnetization vector evolving in the temperature-dependent energy potential given by equation (1) for zero external field ($\boldsymbol{B} = 0$). Simulations are conducted in 0.1 K steps over the temperature range of 295-325 K, with critical SRT temperatures set at $T_L = 302$ K and $T_U = 322$ K.

For each simulated temperature value, an independent MC simulation is carried out. Each MC run begins with the magnetization at its equilibrium angle $\theta_{\min}(T)$, corresponding to the free energy minimum. In cases where multiple degenerate minima exist, the minimum located within the quadrant [0°, 90°] is selected as the initial state. The simulation proceeds for $n = 7000$ MC sweeps, with the magnetization undergoing random angular fluctuations $\delta\theta$ drawn from a uniform distribution in the range $[-3 \text{ rad}, +3 \text{ rad}]$.

The updated magnetization angle $\theta_{n+1} = \theta_n + \delta\theta$ is accepted if the Monte Carlo condition [35] is fulfilled, i.e. $\theta_{n+1}$ either reduces the free energy or:

$$e^{\left(-\frac{(F_{n+1}-F_n)}{F_{\text{therm}}}\right)} > x$$

(7)

Here, $x$ is a random variable drawn from a uniform distribution in $[0,1]$. $F_{\text{therm}}$ represents the thermal free energy. Note that because we will only regard normalized values of the free energy and simulated MC trajectories in this manuscript, hence $F_{\text{therm}}$ is considered as a dimensionless constant.

### III. RESULTS AND DISCUSSION

We perform FemNoC measurements on $Sm_{0.7}Er_{0.3}FeO_3$ across various out-of-plane magnetic fields and temperatures within the SRT region, two measurement approaches are employed:

1. Temperature scan: The sample is held at a constant magnetic field while FemNoC measurements are taken at discrete temperatures ranging from 293 K to 325 K. After completing each temperature scan, the sample is cooled back to 293 K to reset it to a reproducible magnetic state. Subsequently, the magnetic field is ramped to the next value within the range of -350 mT to +350 mT before initiating the next temperature scan sequence.

2. Field scan: FemNoC and static magnetization measurements are performed at a fixed sample temperature while sweeping the external magnetic field from +300 mT to -300 mT and back. The temperature is then increased by 2 K prior to the next scan, covering a total temperature range from 296.15 K to 308.15 K.

*Contact author: marvin.weiss@uni-konstanz.de

## A. Temperature scan results

FIG 2a presents the FemNoC correlation waveforms recorded in temperature scans. These traces are symmetric around $\Delta t = 0$, characteristic of an autocorrelation function, with minor asymmetries attributed to background noise. The waveform shape exhibits strong dependence on temperature and external field. Especially, a pronounced increase in the time-zero correlation amplitude (variance) is observed at higher temperatures. This increase is suppressed under applied magnetic fields.

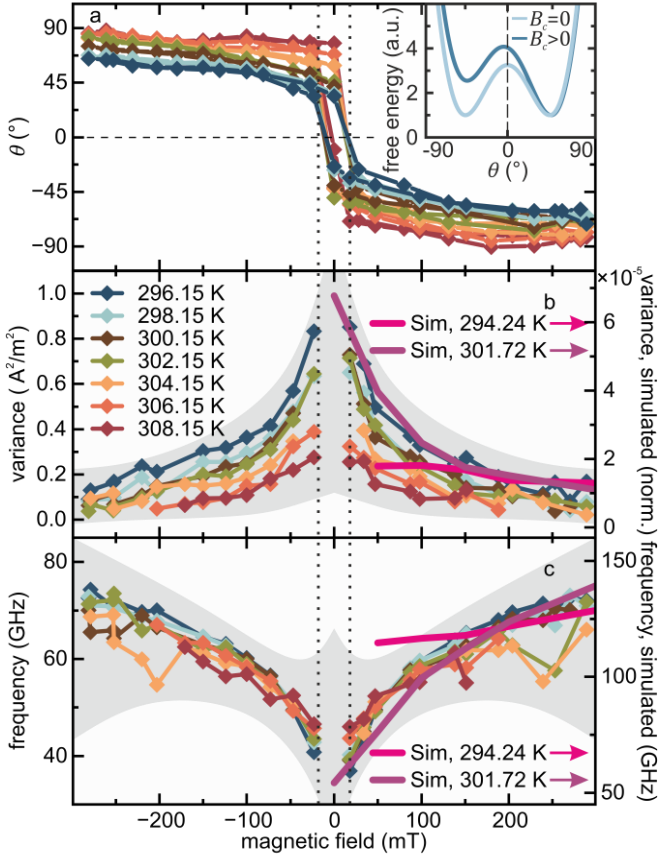

FIG 5: Field dependence of the experimentally determined and simulated $c$-axis component of static magnetization and quasi-ferromagnetic (qF) magnon noise dynamics in $Sm_{0.7}Er_{0.3}FeO_3$. **a** Experimentally determined net magnetization angle $\theta$ as a function of the external field. Inset: Free energy calculated from Eq. (1) as a function of $\theta$ without (light blue curve) and with external magnetic field (dark blue curve) along the crystalline $c$-axis in $Sm_{0.7}Er_{0.3}FeO_3$ held at $T = 305$ K. **b,c** Variance and center frequency of the qF magnon fluctuations as a function of the external magnetic field along the crystalline $c$-axis. The gray shaded area represents the uncertainty of the extracted values, while the dotted vertical lines mark the field range where pronounced random telegraph noise occurs. RTN leads to increased uncertainty, which is why we omit the data in this region. Details on the uncertainty evaluation are provided in Appendix B.

Two distinct dynamical features are evident in the data: A damped picosecond oscillation around the zero-crossing line, present in all waveforms (see FIG 2a). This oscillation is identified as fluctuations of the quasi-ferromagnetic (qF) magnon mode [9,11]. An additional exponential decay lasting tens of picoseconds is detected at -1 mT for 300.15 K and 303.15 K. This decay is indicative of picosecond spontaneous switching between two energetically degenerate quasi-equilibrium states (random telegraph noise, RTN) [9]. These observations are reproduced in atomistic spin model simulations (Methods C), as shown in FIG 2b, which capture the temperature and field dependence of the noise variance and temporal dynamics.

In addition to fluctuation dynamics, our FemNoC setup also resolves the static out-of-plane magnetization component $\boldsymbol{M}$, allowing extraction of its angle $\theta$ relative to the crystalline $a$-axis (Methods B). The temperature dependence of $\theta$ is shown in FIG 3a. For external fields below 60 mT and temperatures $T < 299$ K, $\theta \approx 0°$ (i.e. $\boldsymbol{M}||a$). Above 299 K, $\theta$ increases progressively, saturating at 90° for $T > 321$ K ($\boldsymbol{M}||c$), characteristic of an SRT [36–38]. This finding allows us to identify the critical temperatures as $T_L \approx 299$ K and $T_U \approx 321$ K in the low-field regime.

For external fields exceeding 60 mT, $\theta$ deviates from zero even at the lowest temperatures, indicating a field-induced deflection of $\boldsymbol{M}$ towards the $c$-axis. This shifts the SRT onset to lower temperatures, consistent with previous observations in $ErFeO_3$ [39]. Note that in a Landau-type free energy description, the SRT consists of two seconder-order phase transitions at $T_L$ and $T_U$, with order parameters $M_c = \cos(\theta)$ and $M_a = \sin(\theta)$, respectively. In this picture, application of magnetic fields along the $c$-axis breaks the symmetry (because now $M_c \neq 0$), suppressing critical phenomena at $T_L$ and shifting $T_U$ downward [24,40].

We now turn our discussion to temperature dependence of the fluctuation dynamics. To investigate the individual dynamic features, we fit the correlated noise waveforms using:

$$\langle M_z(t) M_z(t + \Delta t) \rangle = \sum_i A_i e^{-\frac{|\Delta t|}{\tau_i}} \cos(2\pi f_i \Delta t) \tag{8}$$

where $\langle M_z(t) M_z(t + \Delta t) \rangle$ are the correlated out-of-plane magnetization fluctuations as a function of the time delay $\Delta t$, $A_i$ are the noise amplitudes (variances), $\tau_i$ are damping constants, and $f_i$ are the center frequencies. In the experiment, two fluctuation features – RTN and qF magnon noise – are observed, while simulations additionally resolve the quasi-antiferromagnetic (qAF) magnon mode [9]. Consequently, we fit experimental and simulated waveforms using $i = \{RTN, qF\}$ and $i = \{RTN, qF, qAF\}$, respectively (see Appendix A for fitting details).

FIG 3b,c show the temperature dependence of the qF mode variance $A_{qF}$ and resonance frequency $f_{qF}$. For fields below 60 mT, the qF mode variance exhibits a sharp enhancement near $T_L$, followed by a rapid decrease for higher temperatures. This observation is consistent with the magnetization rotation during the SRT [9]. Furthermore, this noise enhancement is accompanied by a softening of the qF mode resonance frequency $f_{qF}$. Note that RTN, with frequencies comparable to $f_{qF}$, emerges in the same temperature range as the qF mode softening. This increases the uncertainty in both qF mode variance and frequency estimations (error bars in FIG 3b,c). Appendix B discusses the uncertainty analysis in detail.

Both the variance enhancement and frequency softening are most pronounced at $B = -1$ mT and diminish as the field increases. For $B > 60$ mT, these effects vanish, indicating field-induced suppression of the critical behavior at $T_L$. This aligns with the expectation that critical fluctuations at a second-order phase transition weaken as external fields modify the free energy landscape.

*Contact author: marvin.weiss@uni-konstanz.de

All these experimental trends are reproduced in atomistic spin noise simulations. The net magnetization rotates from $\theta = 0°$ to $\theta = 90°$, initiating at $T_{L,sim} \approx 302$ K and concluding at $T_{U,sim} \approx 322$ K (FIG 3d). The simulated qF mode variance and resonance frequency (FIG 3e,f) exhibit similar temperature-dependent behavior, including variance enhancement at $T_{L,sim}$ and mode softening at both $T_{L,sim}$ and $T_{U,sim}$. The complete analysis of the atomistic spin noise simulation results, including temperature dependence of the qAF mode and RTN is given in Appendix D.

Note that our FemNoC experiment is only sensitive to fluctuations along the laser propagation direction, which in our setup is aligned with the crystalline $c$-axis. Therefore, we cannot directly resolve qF mode fluctuations near $T_U$ because these take place along the crystalline $a$-axis.

Applying a magnetic field suppresses both variance enhancement and mode softening, albeit less strongly in simulations than in experiments. The discrepancy likely stems from practical limitations in fine-tuning simulation parameters to match experimental details of the reorientation transition (see Methods C). Nonetheless, the overall temperature and field dependence of the qF mode variance and resonance frequency show excellent qualitative agreement between experiment and simulation.

### B. Comparison to Monte Carlo simulation

We now investigate the dependence of thermal fluctuations on the free energy landscape using phenomenological Monte Carlo (MC) simulations. The free energy is computed at multiple temperatures across the SRT, following equation (1). At each temperature, we perform a MC simulation of a magnetization vector subjected to random angular fluctuations within the corresponding potential. Further details on the MC simulations can be found in Methods D. The computed free energies along with simulated MC trajectories are shown in FIG 4a.

Below $T_{L,sim}$, the free energy exhibits a dumbbell shape with minima at $\theta = 0°$ and $\theta = 180°$, indicating that the magnetization easy axis lies along the crystalline $a$-axis. This results in two possible domain configurations with magnetization pointing along either the positive or negative $a$-direction. In this regime, the MC trajectory remains confined to a narrow range around the energy minimum.

At $T = T_{L,sim}$, the free energy landscape softens, characterized by a broadening of the potential well. As a result, the MC trajectory explores a wider range of angles. During the SRT ($T_{L,sim} < T < T_{U,sim}$), the easy axis splits into two branches, rotating in opposite directions as the transition progresses. In this regime, the magnetization can occupy four distinct domain configurations corresponding to the degenerate free energy minima [24,41]. Near $T_{L,sim}$ (e.g., 304.7 K), the energy barrier separating adjacent minima is small compared to the thermal energy, allowing for spontaneous switching between these states. Note that the RTN magnitude in the MC simulation appears significantly larger than observed in FemNoC experiments [12], which is a consequence of the chosen simulation parameters.

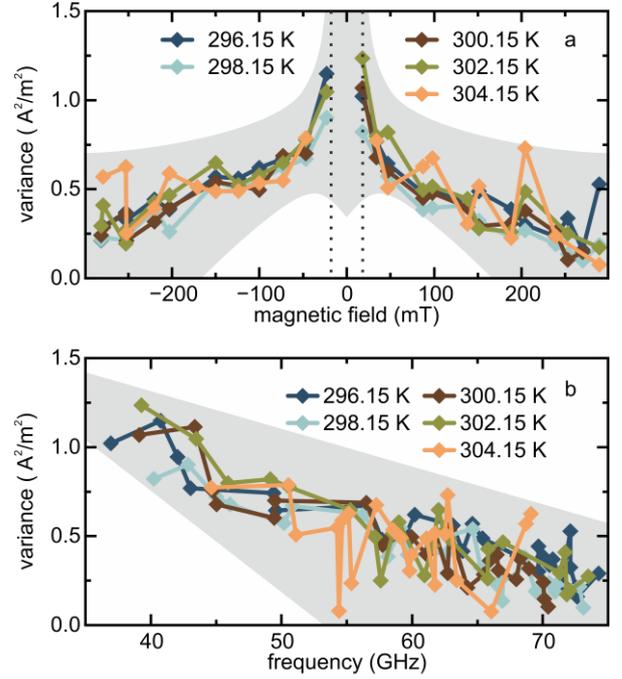

FIG 6: Field and frequency dependence of the SRT compensated quasi-ferromagnetic (qF) magnon noise variance in $Sm_{0.7}Er_{0.3}FeO_3$. **a** Total qF mode variance as a function of the external magnetic field along the crystalline $c$-axis. **b** Total qF mode variance as a function of the qF mode resonance frequency. The gray shaded area represents the uncertainty of the extracted values, while the dotted vertical lines mark the field range where pronounced random telegraph noise occurs. RTN leads to increased uncertainty, which is why we omit the data in this region. Details on the uncertainty evaluation are provided in Appendix B.

At higher temperatures (e.g., 312 K), the energy barriers become too large to be to overcome. As a result, fluctuations remain confined to the local minimum in which the Monte Carlo simulation is initialized. As $T_{U,sim}$ is approached, the two easy axes converge toward the $c$-axis direction. Just below $T_{U,sim}$, the energy barrier between adjacent minima becomes small enough to permit RTN along the crystalline $a$-axis. For $T > T_U$, the two easy axes merge, forming new free energy minima at 90° and 270° (easy axis now parallel to $c$-axis), completing the spin reorientation transition.

FIG 4b shows the variance of the MC trajectories projected along $a$-axis and $c$-axis. The $c$-axis variance exhibits a pronounced enhancement near $T_{L,sim}$, attributed to the softening of the free energy potential. At higher temperatures, the emergence of RTN further amplifies the noise, before eventually vanishing as the SRT progresses, leading to a significant decrease in variance. The $a$-axis variance follows a similar but inverted trend, with a pronounced enhancement near $T_{U,sim}$. These findings closely align with both experimental observations and atomistic simulations, suggesting that the experimentally observed trends can be understood as a direct consequence of the temperature-dependent free energy landscape. This further implies that FemNoC enables direct mapping of the magnetization potential landscape.

*Contact author: marvin.weiss@uni-konstanz.de

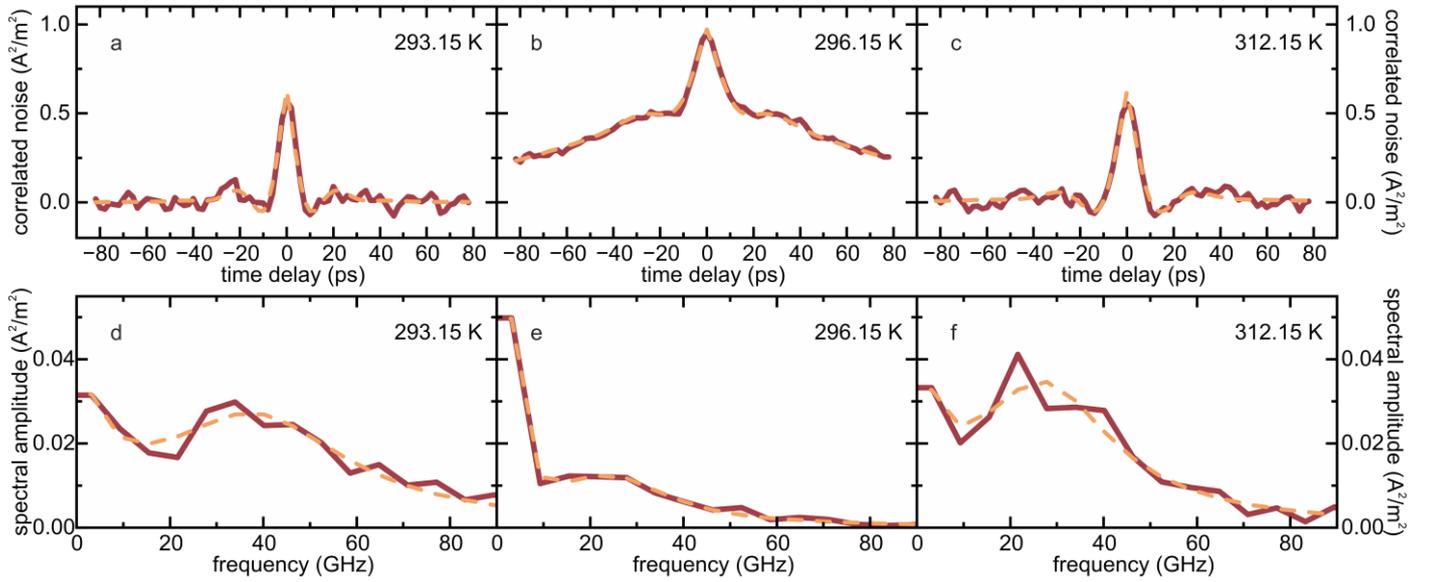

FIG 7: Example of parameter extraction via time-domain fitting of the measured correlation time traces. **a-c** Correlation time traces (red) recorded for the $Sm_{0.7}Er_{0.3}FeO_3$ sample held at an external out-of-plane magnetic field of 5 mT and time-domain fits (orange) using the fitting function $\langle M_c(0)M_c(\Delta t)\rangle = A_{RTN}e^{-\frac{|\Delta t|}{\tau_{RTN}}}\cos(2\pi f_1 \Delta t) + A_{qF}e^{-\frac{|\Delta t|}{\tau_{qF}}}\cos(2\pi f_{qF}\Delta t)$. **d-f** Corresponding frequency spectra of the measured correlation time traces (red) and time-domain fits (orange).

### C. Field scan results

The results of the field scan are presented in FIG 5. FIG 5a shows the net magnetization angle $\theta$ as a function of magnetic field. The data reveals clear hysteretic behavior between up and down sweeps, which varies with temperatures in both saturation angle and steepness. This hysteresis suggests the formation of magnetic domains within the corresponding field range [41].

At $T = 308.15$ K, the field loop exhibits a rectangular shape characteristic of an easy-axis magnetization loop. As the temperature decreases, the loop progressively adopts a hard-axis-like shape. This trend aligns with the known SRT in $Sm_{0.7}Er_{0.3}FeO_3$, where the easy axis rotates from $\boldsymbol{M}||a$ at $T < T_L$ to $\boldsymbol{M}||c$ at $T > T_U$ (see FIG 4). Consequently, the expected field loop behavior transitions from easy-axis-like to hard-axis-like as temperature increases [41], in agreement with our observations.

Interestingly, no hysteresis is expected when the external field is applied along the hard axis [41], which should be the case for $T < T_L$ (see FIG 4). Nevertheless, our lowest recorded temperature (296.15 K) still shows a significant hysteretic effect, despite being below $T_L$ as identified in the temperature scan. This discrepancy is likely due to a slight sample misalignment, such that the external magnetic field is not perfectly parallel to the crystalline easy axis ($a$-axis), leading to finite hysteretic effects.

The inset of FIG 5a illustrates the change in the free energy potential of $Sm_{0.7}Er_{0.3}FeO_3$ under an external magnetic field applied along the crystalline c-axis. At $B_c = 0$, the system exhibits two degenerate minima at $\pm 45°$. The application of a field $B_c > 0$ lifts this degeneracy, rendering the local minimum at $-45°$ energetically less favorable, while the global minimum at $+45°$ is stabilized and stiffened. This behavior reflects a field-induced bias in the domain population: domains with magnetization components parallel to the external field are energetically favored, whereas those with antiparallel components are suppressed, consistent with the observed hysteresis curves.

FIG 5b shows the $c$-axis qF mode variance as a function of magnetic field along the $c$-direction at different temperatures. The noise amplitude systematically decreases with increasing field, coinciding with a characteristic increase in the qF mode frequency (FIG 5c). Due to a significant enhancement of RTN in the field range of -22 mT to +17 mT, we omitted data points in this interval (see Appendix B, FIG 8b). Both the variance reduction and frequency enhancement are reproduced by the atomistic spin noise simulations (purple curves in FIG 5b,c). The qF mode variance also differs between temperatures. This can partially be attributed to temperature- and field-induced changes in $\theta$, which in turn modify $c$-axis component of the qF mode fluctuations.

To isolate the contribution of $\theta$, we computed the absolute transverse magnetization noise $|\delta\boldsymbol{M}|$ in the $a$-$c$-plane from the $c$-axis projection $\delta M_c$, using the trigonometric relationship:

$$|\delta\boldsymbol{M}| = \frac{\delta M_c}{\cos(\theta)} \tag{9}$$

The total magnetization fluctuation $|\delta\boldsymbol{M}|$ as a function of magnetic field is shown in FIG 6a. Note that the 306.15 K and 308.15 K measurements are omitted here, due to the high uncertainty associated with the small noise amplitude. Even after correcting for the SRT, a negative trend remains, indicating a suppression of spin fluctuations with increasing field. Simultaneously, the relationship between noise variance and qF mode frequency persists, as shown in FIG 6b. This further corroborates the notion: A magnetic field applied along the easy axis stiffens the free energy potential (see inset of FIG 5a), reducing spin fluctuations, while increasing the qF mode frequency.

*Contact author: marvin.weiss@uni-konstanz.de

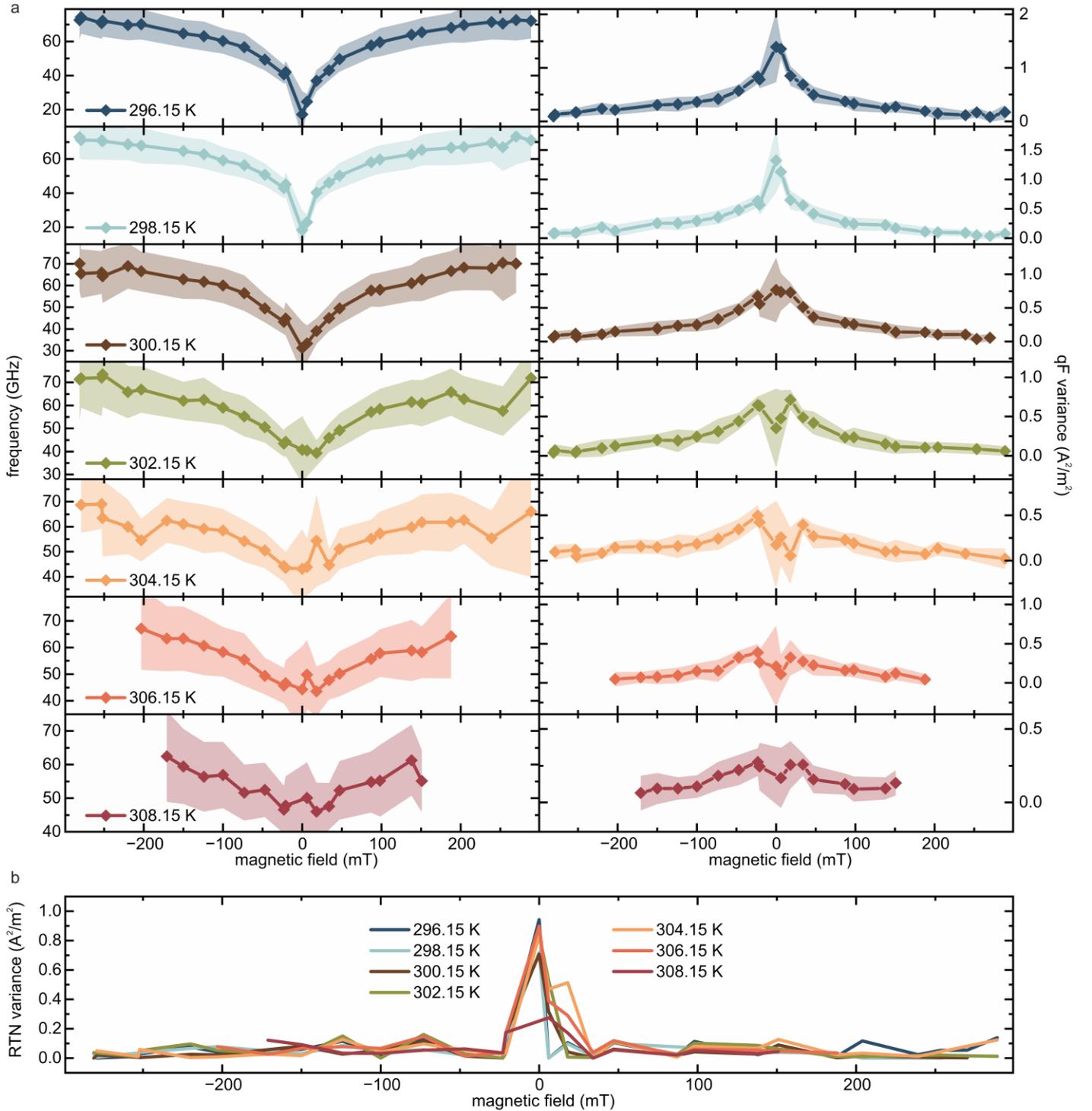

FIG 8: Fitting parameters and corresponding uncertainties obtained from the field scan. **a** Frequency and variances of the qF magnon fluctuations for different sample temperatures as a function of external magnetic field in the out-of-plane direction. **b** RTN variances for different sample temperatures as a function of external magnetic field in the out-of-plane direction.

## IV. CONCLUSION

We investigated ultrafast magnetization fluctuations in the antiferromagnet $Sm_{0.7}Er_{0.3}FeO_3$ using femtosecond noise correlation spectroscopy (FemNoC). By systematically varying temperature across the spin reorientation transition and applying external magnetic fields, we explored the relationship between spin fluctuations, the free energy landscape, and quasi-ferromagnetic (qF) magnon properties.

Our findings show that the magnitude of spin fluctuations at picosecond timescales is dictated by the free energy potential, with fluctuations increasing in regions where the energy landscape softens. Monte Carlo and atomistic spin noise simulations confirm this trend, highlighting a direct correlation between fluctuation amplitude and variations in the free energy curvature. Furthermore, we demonstrate that an external magnetic field suppresses spin fluctuations while increasing the qF magnon frequency, an effect attributed to field-induced stiffening of the free energy potential. These trends are consistently observed in both experiment and simulation.

Overall, our results provide insight into the tunability of ultrafast spin fluctuations through external fields, which is relevant for the development of high-frequency, low-noise spintronic devices. Additionally, they underscore the potential of FemNoC as a tool for characterizing magnetization potentials in complex magnetic systems.

*Contact author: marvin.weiss@uni-konstanz.de

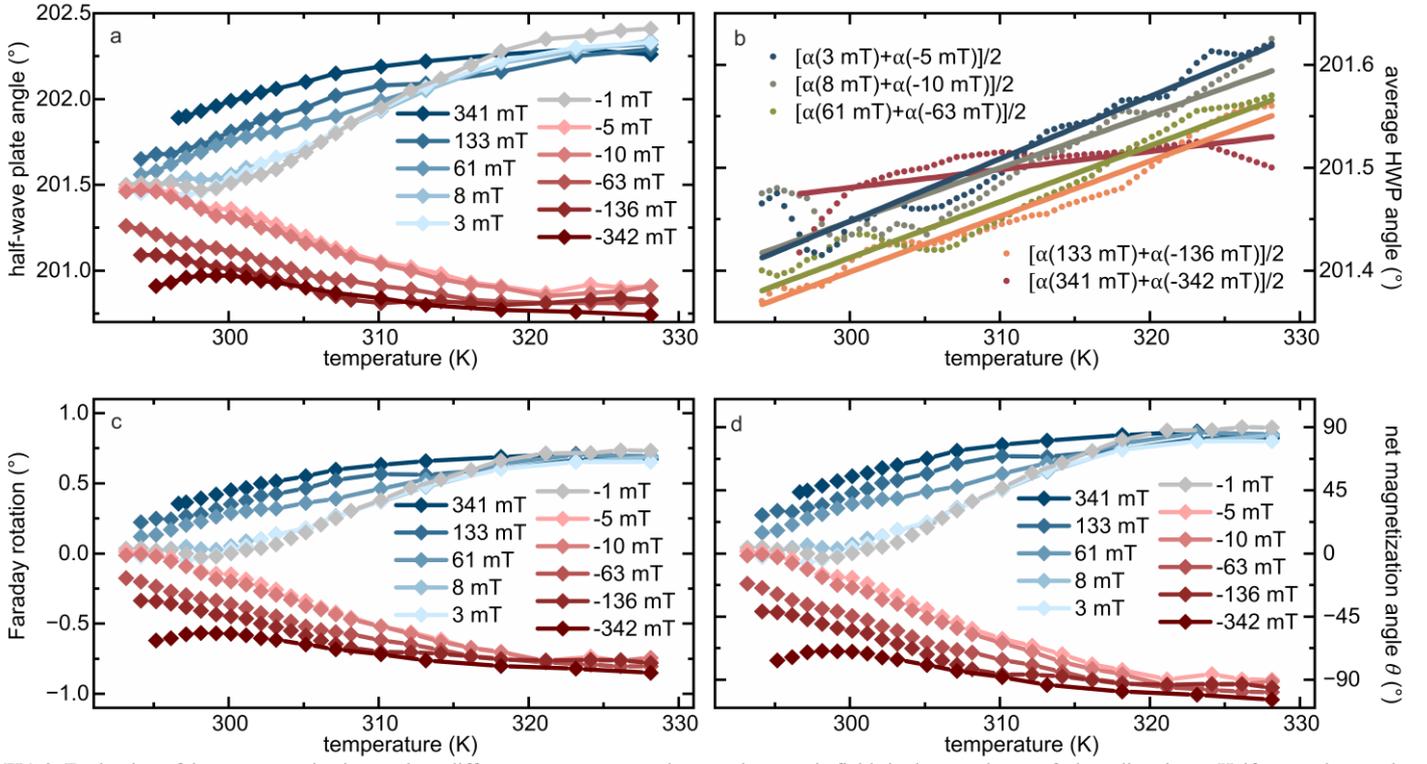

FIG 9: Evaluation of the net magnetization angle at different temperatures and external magnetic fields in the sample out-of-plane direction. **a** Half-wave plate angle (raw data) as a function of temperature. **b** Interpolated average half-wave plate (HWP) angle of field pairs of similar absolute magnitude as a function of temperature (dotted lines) and linear fits (solid lines). **c** Offset and asymmetric background corrected Faraday rotation as a function of temperature. **d** Temperature dependence of the angle $\theta$ between net magnetization and crystalline $a$-axis in $Sm_{0.7}Er_{0.3}FeO_3$. Note that the $\theta$ curves for $B = -63$ mT, $-5$ mT, $3$ mT, $61$ mT are taken from reference [12].

### ACKNOWLEDGEMENTS

We thank the Scientific Engineering Services of the University of Konstanz for fabricating the custom-built electromagnet. We also thank Dr. Jakob Holder for his valuable feedback and guidance on the Monte Carlo simulation. This research was supported by the Japan Society for the Promotion of Science (JSPS) KAKENHI (Grant Nos. JP21K14550, JP24H00317, JP24H02232, and JP23K17748), and the Deutsche Forschungsgemeinschaft (DFG, German Research Foundation) - Project No. 425217212-SFB 1432. The atomistic spin noise simulations were performed on SCCKN, the high-performance computer cluster at the University of Konstanz.

### COMPETING INTERESTS

The authors declare no competing interests.

### DATA AVAILABILITY

The datasets generated in the present study are available from the corresponding author on reasonable request.

*Contact author: marvin.weiss@uni-konstanz.de

## APPENDIX A: POST PROCESSING AND ANALYSIS OF EXPERIMENTAL DATA:

The experimental data presented in this manuscript is subject to several post-processing and analysis steps. First, background correlations are eliminated using a linear subtraction procedure. Here, we define the background as a linear function connecting the average values of the first and last 20 ps of the respective correlation waveform. In the next step, this background function is linearly subtracted from the raw data. Note that the correlation of random telegraph noise (RTN) is an exponentially decaying function with the exponential being proportional to the mean dwell time of the RTN process [9]. Consequently, the exponential decay does not settle to zero within our measurement time window for slow RTN processes (long dwell times). To retain this finite y-axis offset and avoid artifacts from the linear background subtraction, all correlation waveforms showing significant RTN contributions, e.g., FIG 7b, are subtracted by the average background of all other waveforms.

To express our FemNoC data in absolute magnetization units of $(A/m)^2$, we calibrate the waveforms using the protocols described in ref. [12]. The calibration factors used in this manuscript must be adjusted according to the optical probe powers employed during the measurements. Specifically, the raw data for the temperature scans (FIG 2 and FIG 3) and field scans (FIG 5 and FIG 6) were acquired at optical powers of 0.95 mW and 1.09 mW per probe beam, respectively. This results in calibration factors of 8.7 $\frac{\mu V}{(A/m)^2}$ and 11.4 $\frac{\mu V}{(A/m)^2}$, respectively, which scale quadratically with optical power, as detailed in ref. [12]. The raw voltage data from the temperature and field scans are divided by these calibration factors to yield the correlated noise in units of $(A/m)^2$. For details on the uncertainty analysis of the calibration procedure, please refer to ref. [12]. Representative calibrated correlation waveforms are shown in FIG 2 and FIG 7.

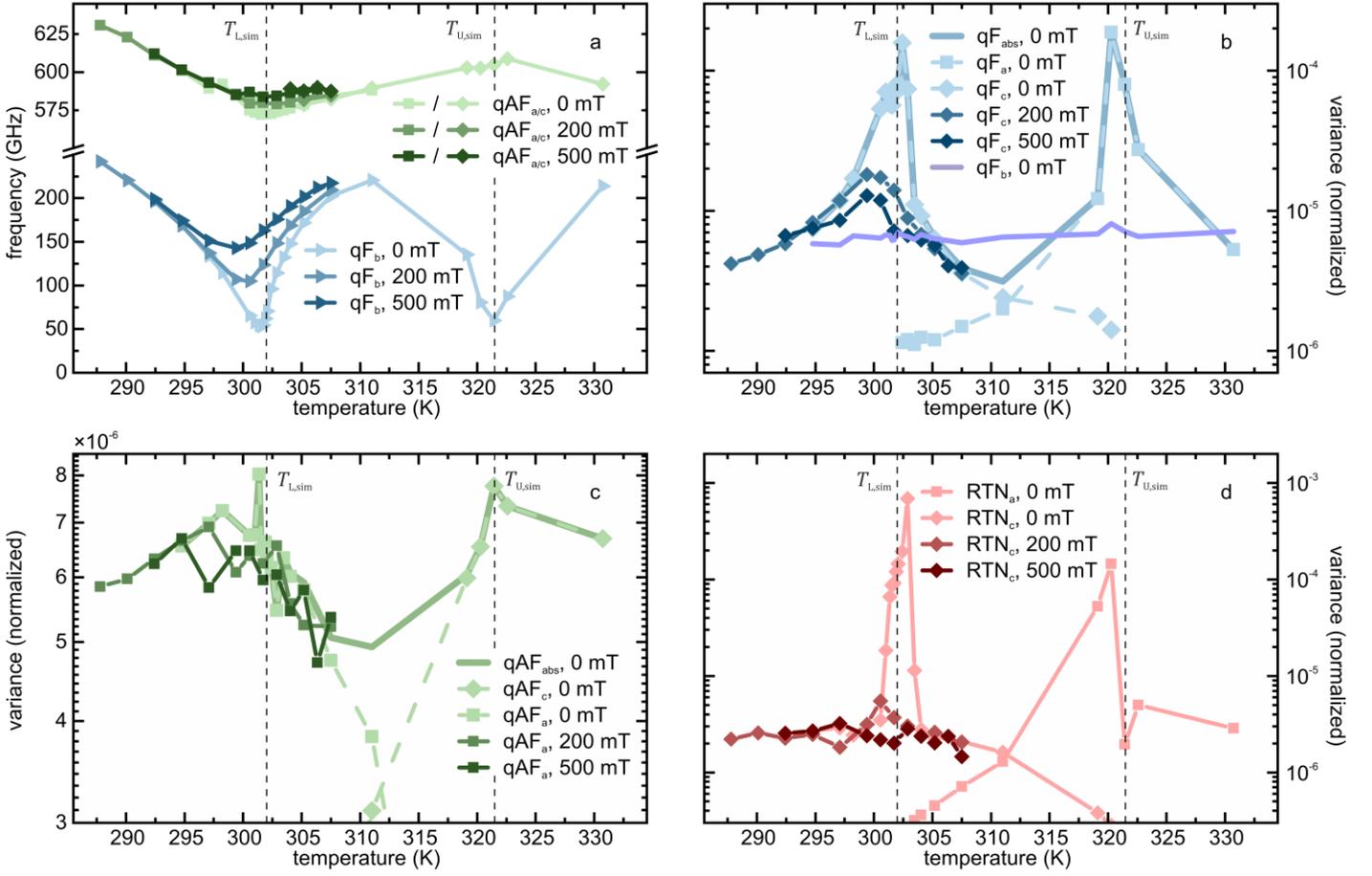

FIG 10: Frequencies and variances of simulated spin noise in $Sm_{0.7}Er_{0.3}FeO_3$ along different crystallographic directions. **a** Frequency of the quasi-ferromagnetic (qF, blue) and quasi-antiferromagnetic (green, qAF) magnon modes along $a$-axis (squares), $b$-axis (triangles), and $c$-axis (diamonds) as a function of temperature. The dotted vertical lines indicate the lower and upper critical temperatures $T_L$ and $T_U$ of the spin reorientation transition. **b** qF mode variances as a function of temperature along $a$-axis (blue squares), $b$-axis (purple line), and $c$-axis (blue diamonds). The blue solid line represents the absolute values (abs) of $a$- and $c$-axis data points. **c** qAF mode variances as a function of temperature along $a$-axis (green squares) and $c$-axis (green diamonds). **d** Variances of the random telegraph fluctuations along $a$-axis (red squares) and $c$-axis (red diamonds).

Analysis of the correlated noise of both experimental and simulated correlation waveforms is performed with the fitting function given in equation (8). Examples of the time-domain fitting of experimental data and the corresponding Fourier spectra of data and fits are shown in FIG 7.

Note that when the qF mode and RTN noise features show finite overlap in the frequency domain, their amplitudes, damping, and frequencies can no longer be unambiguously attributed to a single source. This effect needs to be considered in the analysis, and corresponding uncertainties must be attributed to the fitting parameters. This is further discussed in Appendix B.

## APPENDIX B: UNCERTAINTY EVALUATION OF EXPERIMENTAL PARAMETERS

The experimental values presented in FIG 3, FIG 5, and FIG 6 exhibit finite uncertainties, primarily due to limited signal strength and the time-domain fitting procedure (see Appendix A). The latter effect becomes particularly relevant when the individual noise features exhibit spectral overlap, such as when RTN is comparable in magnitude to the magnon noise. To account for these effects, we estimate the uncertainties of our data using a combination of absolute, relative, and RTN amplitude-dependent contributions. The resulting uncertainty functions are summarized in Table 1. FIG 8a presents the frequencies and variances of the qF magnon fluctuations obtained from the field scan (FIG 5), along with their calculated uncertainties. The corresponding RTN variances are shown in FIG 8b.

As discussed in the main manuscript, applying an external magnetic field alters the potential landscape, leading to a strong reduction in qF mode variances. This decrease in signal-to-noise increases the uncertainty in the qF mode frequency $f_{qF}$. Conversely, at low external fields, the qF magnon noise is enhanced, coinciding with the emergence of RTN at similar frequencies (FIG 7b,e). The resulting spectral overlap significantly increases the uncertainty in both $A_{qF}$ and $f_{qF}$. Consequently, we exclude data points from the field scan results for small external fields ($B \in [-22\,\text{mT}, +17\,\text{mT}]$). The same uncertainty analysis is applied to the temperature scan data presented in FIG 3.

Table 1: Definition of uncertainty functions $u$ corresponding to the fitting parameters extracted from experimental correlation wavefunctions. $A_{RTN}$ and $A_{qF}$: variances of random telegraph noise and qF-mode; $f_{qF}$: frequency of qF-mode; $S = 0.05\left(\frac{A}{m}\right)^2$: Magnetization sensitivity of the experimental setup estimated from the background noise.

*Contact author: marvin.weiss@uni-konstanz.de

| Fit parameter | Uncertainty function |
|---|---|
| $A_{\text{RTN}}$ | $u(A_{\text{RTN}}) = \underbrace{S}_{\substack{\text{absolute error}\\\text{(sensitivity)}}} + \underbrace{0.3 \cdot A_{\text{RTN}}}_{\substack{\text{relative error}\\\text{(fit accuracy)}}} + \underbrace{0.1 \left(\frac{\text{A}}{\text{m}}\right)^2}_{\substack{\text{absolute error}\\\text{(dc offset)}}}$ |
| $A_{\text{qF}}$ | $u(A_{\text{qF}}) = \underbrace{S}_{\substack{\text{absolute error}\\\text{(sensitivity)}}} + \underbrace{0.1 \cdot A_{\text{qF}}}_{\substack{\text{relative error}\\\text{(fit accuracy)}}} + \underbrace{0.5 \cdot A_{\text{RTN}}}_{\text{RTN amp. dependent error}}$ |
| $f_{\text{qF}}$ | $u(f_{\text{qF}}) = \underbrace{3 \text{ GHz}}_{\text{absolute error}} + \underbrace{0.05 \cdot f_{\text{qF}}}_{\substack{\text{relative error}\\\text{(fit accuracy)}}} + \underbrace{8 \text{ GHz} \cdot \frac{S}{A_{\text{qF}}}}_{\text{qF amp. dependent err.}} + \underbrace{0.5 \text{ GHz} \cdot \frac{A_{\text{RTN}}}{S}}_{\text{RTN dependent err.}}$ |

## APPENDIX C: EVALUATION OF THE NET MAGNETIZATION ANGLE

As discussed in Methods B, our FemNoC setup also allows us to measure the net magnetization angle $\theta$ relative to the crystallographic $a$-axis (in-plane direction) of $Sm_{0.7}Er_{0.3}FeO_3$. This is achieved by determining the half-wave plate angles at which the BPDs are balanced. This is the case, when incident light on the Wollaston prisms is polarized at 45° relative to their optical axes. Below the lower threshold temperature of the spin-reorientation transition $T_L$, the sample's net magnetization lies along the $a$-axis, meaning no static magnetization component exists along the laser's propagation direction. Above $T_L$, the magnetization rotates by an angle $\theta(T)$ towards the $c$-axis, acquiring a finite out-of-plane component. This component induces a static Faraday rotation in the optical probes, which must be compensated by adjusting the half-wave plates before each FemNoC measurement. Consequently, the half-wave plate angles encode information about $\theta$.

FIG 9a displays the absolute half-wave plate angles as a function of sample temperature under different out-of-plane magnetic fields. These angles vary significantly with sample temperature and external magnetic field due to the spin-reorientation transition. Notably, slight discrepancies appear between measurements performed at positive and negative fields of similar magnitude. This effect may be explained by temperature-dependent birefringence of the sample, which introduces an additional polarization rotation component that is independent of the sample's magnetization [42,12].

To isolate this asymmetric background, we interpolate the data in FIG 9a and sum two half-wave plate curves measured at positive and negative fields of similar magnitude. FIG 9b presents the results, where linear fits reveal the asymmetric backgrounds. After subtracting the backgrounds from the raw data (FIG 9a), we define the zero-reference level as the average of the values recorded below 297 K for fields $|B| <$ 15 mT. The resulting relative Faraday rotation is shown in FIG 9c.

For temperatures $T > T_U = 321$ K, the Faraday rotation curves stabilize at comparable relative values. Assuming $\theta(T > T_U) = \pm 90°$, we normalize the Faraday rotation data using the maximum (minimum) values of the $-1$ mT ($-5$ mT) curves, along with the previously established zero reference. The extracted net magnetization angles $\theta(T)$ are shown in FIG 9d and, for positive fields only, in FIG 3c. The same reference levels are applied to half-wave plate angles from the field scan, yielding the $\theta$ reference levels obtained in this procedure are also applied to the half-wave plate angles measured in the field scans, resulting in the net magnetization curves shown in FIG 5c.

## APPENDIX D: FULL ANALYSIS OF SIMULATION DATA

FIG 10 represents the complete results of the atomistic simulations. The temperature dependence of the qF and qAF magnon frequencies is shown in FIG 10a. The qF mode exhibits characteristic softening near the critical temperatures, an effect that is suppressed under applied magnetic fields. On the contrary, the qAF mode frequency remains largely temperature-independent, with only a slight field-induced increase.

FIG 10b displays the qF mode variance temperature-dependence for different axis projections. The $c$-axis projection initially increases around $T_L$, reflecting the softening of the free energy potential, before decreasing due to the SRT. Simultaneously, the $a$-axis projection steadily rises, peaking near $T_U$. This noise enhancement is significantly reduced by external magnetic fields, which introduces stiffening of the magnetic potential. In contrast, the $b$-axis projection remains largely unchanged throughout the SRT, suggesting a stable free energy landscape along this direction.

The fluctuation amplitude of the qAF is shown in FIG 10c. While the latter also exhibits an increase near the critical temperatures, this effect is much weaker than for the qF mode. Interestingly, external fields have little impact on the qAF mode noise increase near $T_{L,\text{sim}}$.

Finally, FIG 10d presents the RTN amplitude as a function of temperature and applied magnetic field. RTN appears along the $c$-axis near $T_{L,\text{sim}}$ and along the $a$-axis near $T_{U,\text{sim}}$, in agreement with the Monte Carlo simulations. The application of external magnetic fields strongly suppresses RTN, which we attribute to asymmetric shifting of the potential minima and the associated increase in the effective potential barrier.

*Contact author: marvin.weiss@uni-konstanz.de

*Contact author: marvin.weiss@uni-konstanz.de

*Contact author: marvin.weiss@uni-konstanz.de